\documentclass[
aip,
sd,
amsmath,amssymb,
reprint
]{revtex4-1}

\usepackage{graphicx}% Include figure files
\usepackage{epstopdf,soul}
\usepackage{dcolumn}% Align table columns on decimal point
\usepackage{bm}
\begin{document}
\preprint{AIP/123-QED}
%\preprint{APS/PRE}
\title{Chimera-like behavior in a heterogeneous Kuramoto model: the interplay between the attractive and repulsive coupling}
% Force line breaks with \\
	\author{Nikita Frolov}\email{n.frolov@innopolis.ru}
	\affiliation{Neuroscience and Cognitive Technology Laboratory, Center for Technologies in Robotics and Mechatronics Components, Innopolis University, 420500, Innopolis, The Republic of Tatarstan, Russia}
	
	\author{Vladimir Maksimenko}\email{v.maksimenko@innopolis.ru}
		\affiliation{Neuroscience and Cognitive Technology Laboratory, Center for Technologies in Robotics and Mechatronics Components, Innopolis University, 420500, Innopolis, The Republic of Tatarstan, Russia}
	
	\author{Soumen Majhi}\email{soumen.majhi91@gmail.com}
		\affiliation{Physics and Applied Mathematics Unit, Indian Statistical Institute, 203 B. T. Road, Kolkata-700108, India}

    \author{Sarbendu Rakshit}\email{sarbendu.math@gmail.com}
    		\affiliation{Physics and Applied Mathematics Unit, Indian Statistical Institute, 203 B. T. Road, Kolkata-700108, India}
	
	\author{Dibakar Ghosh}\email{diba.ghosh@gmail.com}
	\affiliation{Physics and Applied Mathematics Unit, Indian Statistical Institute, 203 B. T. Road, Kolkata-700108, India}
	
	\author{Alexander Hramov}\email{a.hramov@innopolis.ru}
			\affiliation{Neuroscience and Cognitive Technology Laboratory, Center for Technologies in Robotics and Mechatronics Components, Innopolis University, 420500, Innopolis, The Republic of Tatarstan, Russia}
			
	\date{\today}
	
%\begin{keywords}
%chimera \sep complex network \sep heterogeneity \sep Kuramoto model \sep explosive synchronization
%\end{keywords}	

	\begin{abstract}
		Interaction within an ensemble of coupled nonlinear oscillators induces a variety of collective behaviors. One of the most fascinating is a chimera state which manifests the coexistence of spatially distinct populations of coherent and incoherent elements. Understanding of the emergent chimera behavior in controlled experiments or real systems requires a focus on the consideration of heterogeneous network models. In this study, we explore the transitions in a heterogeneous Kuramoto model under the monotonical increase of the coupling strength and specifically find that this system exhibits a frequency-modulated chimera-like pattern during the explosive transition to synchronization. We demonstrate that this specific dynamical regime originates from the interplay between (the evolved) attractively and repulsively coupled subpopulations. We also show that the above mentioned chimera-like state is induced under weakly non-local, small-world and sparse scale-free coupling and suppressed in globally coupled, strongly rewired and dense scale-free networks due to the emergence of the large-scale connections.

	\end{abstract}
	
	\maketitle
\begin{quotation}
{\bf Synchronization phenomena in populations of interacting elements are the subject of extensive research in biological, chemical, physical and social systems. 
The process of synchronization refers to the adjustment of rhythms of interacting oscillatory systems, whereas chimera states are characterized by the fascinating coexistence of coherent and incoherent sub-populations  in networks of coupled oscillators.
On another note, discontinuous or explosive transitions to coherence in networks are receiving growing attention these days. The paradigmatic Kuramoto model being able to provide the most effective approach to explain how synchronous behavior emerges in complex systems, there exists significant attempts in exploring both chimera states and explosive transition to synchrony. But, in most of the studies, these two processes have been studied exclusively, without paying attention to a possibility in linking them.  
In contrast to approaches solely concentrating on abrupt transitions to synchrony and the associated hysteresis, we here put forward the emergence of chimera-like behavior on the route to an explosive transition in networks of coupled Kuramoto phase oscillators. 
Complex systems naturally display heterogeneity in its constituents, so in this article, we consider a heterogeneous Kuramoto model and report a frequency-modulated chimera-like pattern during discontinuous transitions to coherence. We reveal that this chimera-like behavior appears due to a coexistence of evolved (not induced) attractively and repulsively coupled populations of oscillators. We further establish that the uncovered type of chimera-like state is excited under weakly non-local, small-world and sparse scale-free coupling and suppressed in globally coupled, strongly rewired and dense scale-free networks. }
\end{quotation}

\section{Introduction}
Network science provides a universal language to create relevant models and understand the behavior of complex systems~\cite{boccaletti2006complex}. Among diverse dynamical phenomena, i.e., synchronization, adaptation, clustering, etc. performed by the complex network models,  \textit{chimera state} is one of the most intriguing types of collective behavior. Originally, it implies the coexistence of coherent and incoherent populations in a symmetrically coupled ensemble of identical nonlinear oscillators \cite{abrams2004chimera}.

For almost two decades from its discovery~\cite{kuramoto2002coexistence}, many aspects of this specific dynamical regime were explored in detail. Specifically, chimera patterns were demonstrated to be a universal phenomenon for the models of different nature, including phase oscillators~\cite{panaggio2015chimera}, oscillators with inertia~\cite{olmi2015chimera,jaros2015chimera}, chaotic systems~\cite{omelchenko2011loss,bogomolov2017mechanisms}, biological neurons based on the Hodgkin-Huxley~\cite{andreev2019chimera}, FitzHugh-Nagumo~\cite{omelchenko2015robustness,shepelev2017new,guo2018spiral}, Hindmarch-Rose \cite{hizanidis2014chimera,bera2016chimera} models. Several remarkable fundamental effects such as coherence-resonance chimera~\cite{semenova2016coherence} and virtual chimera \cite{larger2013virtual,larger2015laser} were discovered in the last few years. Chimeras were also shown to be robust against the topology and reported in globally coupled~\cite{yeldesbay2014chimeralike}, hierarchical~\cite{ulonska2016chimera}, scale-free~\cite{zhu2014chimera} and small-world networks~\cite{rothkegel2014irregular,hizanidis2016chimera}, multilayer~\cite{maksimenko2016excitation,ghosh2016emergence,ghosh2018non,frolov2018macroscopic} and multiscale networks~\cite{makarov2019multiscale}, and even hypergraphs~\cite{bera2019spike}. For a long time observed only in the model systems, chimera patterns were experimentally verified in the mechanical~\cite{kapitaniak2014imperfect,wojewoda2016smallest}, chemical~\cite{tinsley2012chimera}, and optical~\cite{hagerstrom2012experimental} setups.

The chimera behavior is still closely studied as it fits the dynamics of various real-life systems, i.e., social~\cite{gonzalez2014localized} and biological~\cite{hizanidis2015chimera,dutta2015spatial,banerjee2016chimera,kundu2018diffusion,dana2019chimera} systems, power grids~\cite{motter2013spontaneous,pecora2014cluster} etc. Special interest is paid to the application of chimeras in neuroscience~\cite{majhi2019chimera}, since spatio-temporal coherence is a cornerstone of the normal and pathological brain activity~\cite{fries2015rhythms,uhlhaas2006neural}. Earlier, chimera patterns were observed in animals' neural networks~\cite{hizanidis2016chimera,santos2017chimera}. In humans, such forms of the brain activity as epileptic seizures~\cite{andrzejak2016all}, Parkinson's and Alzheimer's disease~\cite{protachevicz2019bistable,coninck2020network}, bump states~\cite{roxin2005role,laing2011fronts}, cognitive functions and resting-state~\cite{bansal2019cognitive,kang2019two} are shown to perform the pronounced properties of chimera behavior.

However, the approach to more realistic models requires consideration of non-homogeneous ensembles since the condition of elements' identity is hardly fulfilled in the real networks. Several studies addressed the problem of network heterogeneity in the context of chimera behavior. Specifically, bifurcation analysis of the Kuramoto network with heterogeneous intrinsic frequencies was performed by  Laing~\cite{laing2009chimera,laing2009dynamics}. Based on the results of numerical and analytical treatment, the author concluded that chimera is robust to such type of heterogeneity. Nkomo et al.~\cite{nkomo2016chimera} demonstrated the chimera state in the ensemble of heterogeneous Belousov-Zhabotinski oscillators both numerically and experimentally. Several works reported that the chimera state could be induced in the presence of phase-lag heterogeneity~\cite{zhu2013reversed,martens2016chimera,choe2017chimera}. Chimera state was also explored in networks with irregular topology~\cite{majhi2017chimera,li2017chimera}. On the other hand, intense research efforts have also been made in order to study mechanisms that lead to discontinuous or explosive transition to synchrony \cite{gomez2011explosive,zhang2015explosive,kachhvah2019delay,jalan2019inhibition}.

Despite the above discussed extensive studies on chimera behavior, even simply constructed complex networks still hide unexpected aspects of this phenomenon due to heterogeneity of its elements. In this paper, we report the emergence of a frequency modulated chimera-like behavior in a non-homogeneous Kuramoto model during an explosive transition of the networked system to a certain level of coherence. We argue that the uncovered chimera-like behavior occurs in weakly non-local, small-world (SW) and sparse scale-free (SF) coupling. We demonstrate that it originates from the self-organization of the entire ensemble into attractively and repulsively coupled populations.

\section{Mathematical Model}
We consider a network of $N$ number of phase oscillators,  in which the dynamics of each node is represented by the following form of the Kuramoto equation:

\begin{equation}
	\begin{split}
		\dot{\phi}_i & = \omega_i + \lambda R_i \sum_{l=1}^{N} A_{il} \sin (\phi_l - \phi_i ),\\
		R_i &= \frac{1}{k_i} \bigg|\sum_{l=1}^N A_{il} e^{\mathrm{j}\phi_l}\bigg|,
	\end{split}
	\label{eq:KO}
\end{equation}
where $\phi_i$, $\omega_i$ and $k_i$ are the phase, natural frequency and the degree of the $i^{th}$ Kuramoto oscillator respectively, also $\mathrm{j}=\sqrt{-1}$. For further simplicity, let us introduce the notation for the effective frequency of the $i^{th}$ oscillator as $f_i = \dot{\phi}_i$. The parameter $\lambda$ is the overall coupling strength. The matrix $A=[A_{il}]$ is the underlying graph adjacency. In the case of regular and SW coupling, it is generated using the Watts-Strogatz (WS) algorithm with $k$ nearest neighbors (in each side of a one-dimensional ring) and the probability $p$ of adding a shortcut in a given row~\cite{watts1998collective}. The SF adjacency matrix is generated using Barab\'{a}si-Albert (BA) algorithm~\cite{barabasi1999emergence} with the growing parameter $m$. $R_i$ represents the local order parameter and evaluates the degree of coherence in the neighborhood of the $i^{th}$ element. It contributes adiabatically to the coupling term and provides the mechanism for explosive synchronization. The values of $\omega_i$ are uniformly distributed over the range $[\omega_0-\frac{\Delta}{2},\omega_0+\frac{\Delta}{2}]$, where $\omega_0$ is the central frequency and $\Delta$ is the width of the frequency range.

To quantify the network's coherence, we use the averaged global order parameter as
\begin{equation}
R = \frac{1}{N(t_{max}-t_{trans})}\int_{t_{trans}}^{t_{max}}\Big|\sum_{l=1}^N e^{\mathrm{j}\phi_l(t)}\Big|~dt,
	\label{eq:Order}
\end{equation}
where $t_{trans}$ and $t_{max}$ respectively denote the transient time and maximal simulation time. Moreover, we illustrate the collective behavior of the Kuramoto model using the mean effective frequency $\langle f_i \rangle$ defined by time averaging instantaneous effective frequency $f_i(t)$ after the transient process.

\section{Results}
Specifically, we consider the dynamical network (\ref{eq:KO}) consisting of $N=100$ oscillators. The value of the central frequency is fixed at $\omega_0=10$. The network model simulation is conducted using the Runge-Kutta method of order 5(4)~\cite{tsitouras2011runge} implemented in the Differential Equation Solver for Julia programming language~\cite{rackauckas2017differentialequations}. To control the accuracy of the numerical integration, we use the adaptive time-stepping with relative tolerance parameter equal to $10^{-6}$, maximal simulation time $t_{max}=2000$ and transient time $t_{trans}=1500$.

\subsection{Observation of the chimera-like behavior}
\label{ssec:Observation}

\begin{figure}[!t]
\centerline{\includegraphics[scale=0.750]{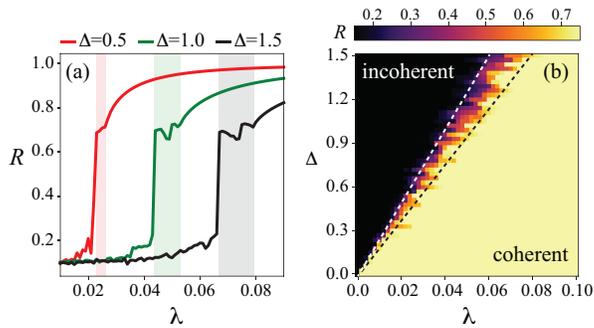}}
\caption{(a) Averaged global order parameter $R$ versus the coupling strength $\lambda$ in the non-locally coupled network of $N=100$ oscillators with $p=0.0$ and $k=10$ for different values of natural frequency distribution width: $\Delta=0.5$ (red); $\Delta=1.0$ (green); $\Delta=1.5$ (black). Shadings highlight the respective areas of partially coherent chimera-like regimes. (b) Phase diagram in the $(\lambda,\Delta)$ parameter plane for the global order parameter $R$, color bar represents its variation.}\label{fig:1}
\end{figure}

Depending on the level of heterogeneity, i.e., the width $\Delta$ of the natural frequency distribution, we observe different transitions to coherence in a Kuramoto model under the adiabatically increasing coupling strength $\lambda$ (Fig~\ref{fig:1}). Obviously, an ensemble with a homogeneous frequency distribution, i.e., for $\Delta=0$, the coupled system \eqref{eq:KO} undergoes a smooth transition to coherence at very small values of the coupling strength. The introduction of heterogeneity in the considered network system (cf. Eq.~(\ref{eq:KO})) leads to the explosive transition to coherence. Here, the incoherence for the values of coupling strength below the critical point $\lambda_{cr}$, is supported by the low degree of local synchrony $R_i$ that reduces the value of coupling term in Eq.~(\ref{eq:KO}). Interestingly, a heterogeneous Kuramoto model does not converge to a global frequency-locking ($\pi$-state) immediately after the explosive transition. Instead, we find a finite-size plateau, where the Kuramoto model exhibits a partially coherent state with the averaged order parameter $R\approx0.7$. As seen in Fig~\ref{fig:1}(a), the way of transition does not depend on the degree of heterogeneity $\Delta$. Notable, that in the case of higher values of $\Delta$, the transition occurs at the greater values of the critical coupling strength $\lambda_{cr}$ and it is followed by a wider `partially coherent' plateau.

To illustrate the effect of the coupling strength $\lambda$ on this chimera-like state for continuous variation of $\Delta$, we plot the global order parameter $R$ in the $(\lambda,\Delta)$ parameter plane in Fig. \ref{fig:1}(b). The region between the dashed white and black lines reflects the existence of chimera-like state. However, the yellow and black regions respectively correspond to the coherent and incoherent states.
The figure explicitly demonstrates the interval of $\lambda$ for which chimera-like state emerges. Interestingly, this interval that supports the chimera-like state improves considerably as $\Delta$ increases. Beyond certain values of the coupling strength $\lambda$ (depending on the width $\Delta$), the coupled Kuramoto oscillators undergoes the coherent state and persists further.

\begin{figure}[!t]
\centerline{\includegraphics[scale=0.75]{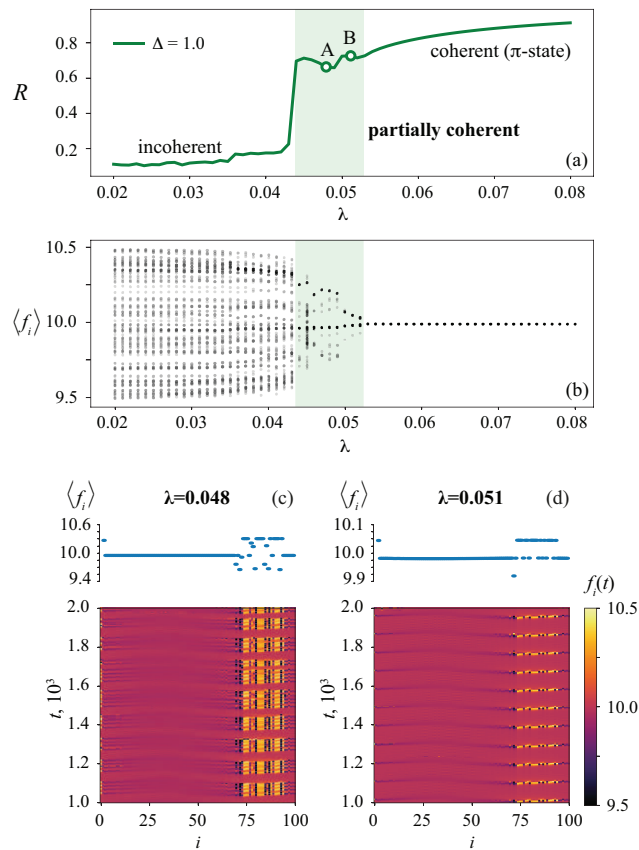}}
\caption{(a) Averaged global order parameter $R$  and (b) the distribution of mean effective frequencies $\langle f_i \rangle$ versus the coupling strength $\lambda$ in the heterogeneous non-locally coupled network ($p=0.0$, $k=10$ and $\Delta=1.0$). Shading highlights the area of partially coherent states. (c,d) Mean effective frequency $\langle f_i \rangle$ profiles (top) and the space-time plots of the instantaneous effective frequency $f_i$ (bottom) for the different values of the coupling strength $\lambda$ corresponding to points A and B: (c) $\lambda=0.048$, point A; (d) $\lambda=0.051$, point B.}\label{fig:2}% Give a unique label
\end{figure}

Let us now take a close look at the transitions in the considered Kuramoto model. Without any loss of generality, we fix $\Delta=1.0$ and consider how the network evolves under the increment of the coupling strength $\lambda$, in terms of the averaged global order parameter $R$ (cf. Fig.~\ref{fig:2}a) and the distribution of mean effective frequencies $\langle f_i \rangle$ (cf. Fig.~\ref{fig:2}b). It is seen that even at $\lambda=0.02$ effective frequencies remain uniformly distributed over the ensemble and are almost unchanged with respect to the initial distribution of natural frequencies, so that $\langle f_i \rangle \approx \omega_i$, $i=1,2,\dots,N$. While coupling strength approaches the critical value of explosive transition $\lambda_{cr}=0.044$, the effective frequencies tend to converge slowly to a central frequency of the initial distribution $\omega_0$. After the critical explosive transition at $\lambda_{cr}=0.044$, a large part of the network elements $N_{coh}$ undergoes the abrupt frequency-locking, so that $\langle f_i \rangle \approx \omega_0$ for all $i \in N_{coh}$. At the same time, a group of oscillators $N_{inc}$ remains desynchronized, i.e., $|\langle f_i \rangle-\omega_0|>>0$ for all $i \in N_{inc}$. Thus, the balance between the heterogeneity of natural frequencies and the coupling strength, which is insufficient to provide a global coherence, supports a partially coherent state in a non-homogeneous Kuramoto ensemble. However, the sharp increase of the network's coherence boosts faster convergence of the remaining part of oscillators to a globally frequency-locked state at  $\lambda_{cr}=0.054$.  

Furthermore, one can see in Fig~\ref{fig:2}a, that the dependency of the global order parameter $R$ on the coupling strength $\lambda$ has two peaks in the area, where the network exhibits partially coherent state. It reflects the switching between two distinct regimes of partial coherence. Let us consider the latter in detail by tracking the network's behavior at points A ($\lambda=0.048$) and B ($\lambda=0.051$) marked with circles in Fig.~\ref{fig:2}a. Figures~\ref{fig:2}c and ~\ref{fig:2}d present the profiles of mean effective frequency $\langle f_i \rangle$ (top row) along with the corresponding space-time plots color-coded by the instantaneous effective frequency $f_i$ (bottom row). We find that both partially coherent states that occurred after the critical transition represent a specific form of a frequency-modulated \textit{`chimera-like'} behavior. Specifically, we observe the coexistence of two distinct clusters: a larger one that is frequency-locked and follow a smooth coherent spatiotemporal profile, however the smaller one evolves in a drifting-like manner. Here, we intentionally refer this regime to as a \textit{`chimera-like'} behavior, since it differs from the classical definition of the \textit{`chimera'} mostly because we here consider a heterogeneous ensemble of phase oscillators. Also, the traditional \textit{chimera} state implies coherence in terms of the phase-locking, instead of the frequency-locking reported here. Despite that, we still observe the relevant feature of \textit{chimera} behavior in the uncovered network dynamics, i.e., the coexistence of spatially dissociated groups of coherent and incoherent network elements, that gives us a fair basis to determine the uncovered phenomenon as a \textit{chimera-like} state. 

Interestingly, the observed chimera-like regimes are not stationary -- the incoherent cluster appears and collapses in time. The way of evolution in time determines the difference between these partially coherent states. The regime at $\lambda=0.048$ formed after the critical transition and presented in Fig.~\ref{fig:2}c is characterized by the fast and irregular burst-like oscillations of the incoherent cluster. On the contrary, an increase of the coupling strength $\lambda$ switches the chimera-like regime to slow and periodic oscillations (Fig.~\ref{fig:2}d).     

\begin{figure}[t]
\centerline{\includegraphics[scale=0.85]{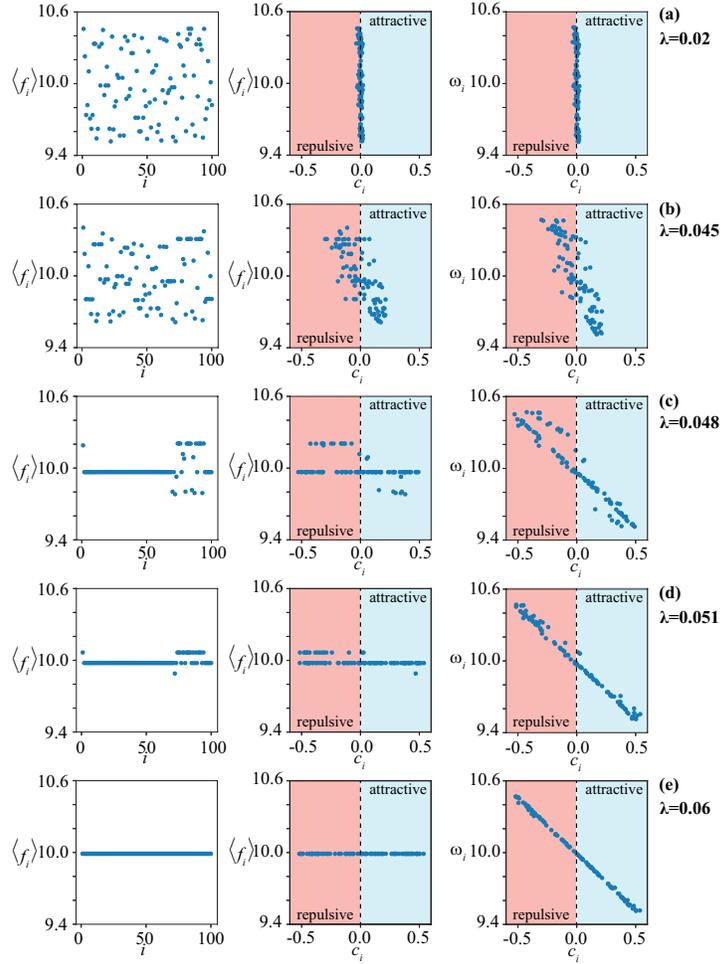}}
\caption{Illustration of the mechanism underlying the chimera-like pattern formation in the non-locally coupled network ($p=0.0$, $k=10$) with heterogeneous natural frequency distribution ($\Delta=1.0$). Averaged effective frequency $\langle f_i \rangle$ profile (left column), its correspondence to the coupling term $c_i$ (middle column) and the natural frequency $\omega_i$ versus the coupling term $c_i$ (right column) for different values of the coupling strength $\lambda$: (a) $\lambda = 0.02$; (b) $\lambda = 0.045$; (c) $\lambda = 0.048$; (d) $\lambda = 0.051$; (e) $\lambda = 0.06$. Blue and red colors highlight the attractive and repulsive coupling areas respectively in the middle and right columns.}\label{fig:3}       % Give a unique label
\end{figure}

\subsection{Birth of chimera-like state: Mechanism}

To understand the mechanism of the birth of chimera-like behavior in a heterogeneous Kuramoto model let us rewrite the model Eq.~(\ref{eq:KO}) in the following form:
\begin{equation}
	\begin{split}
	\dot{\phi}_i & = f_i = \omega_i + c_i,\\
	c_i &= \lambda R_i \sum_{l=1}^{N} A_{il} \sin (\phi_l - \phi_i ),
	\end{split}
\label{eq:KO_mod}
\end{equation}
where we introduce a notion called mean coupling term $c_i$ associated with the $i^{th}$ element's coupling term in the governing Kuramoto equation.

It is clear from the modified Eq.~(\ref{eq:KO_mod}) that frequency-locking $\langle f_i\rangle=\Omega$, where $\Omega$ is a mean-field frequency, implies $\omega_i + \langle c_i \rangle = \Omega$, $i=1, 2, ..., N$. In the case of uniform natural frequency distribution, $\Omega\approx\omega_0$ and, therefore, coupling term should provide the compensation of the difference between the central and natural frequencies of the $i^{th}$ oscillator $\langle c_i \rangle\approx\omega_0 - \omega_i$.

\begin{figure}[!t]
\centerline{\includegraphics[scale=0.75]{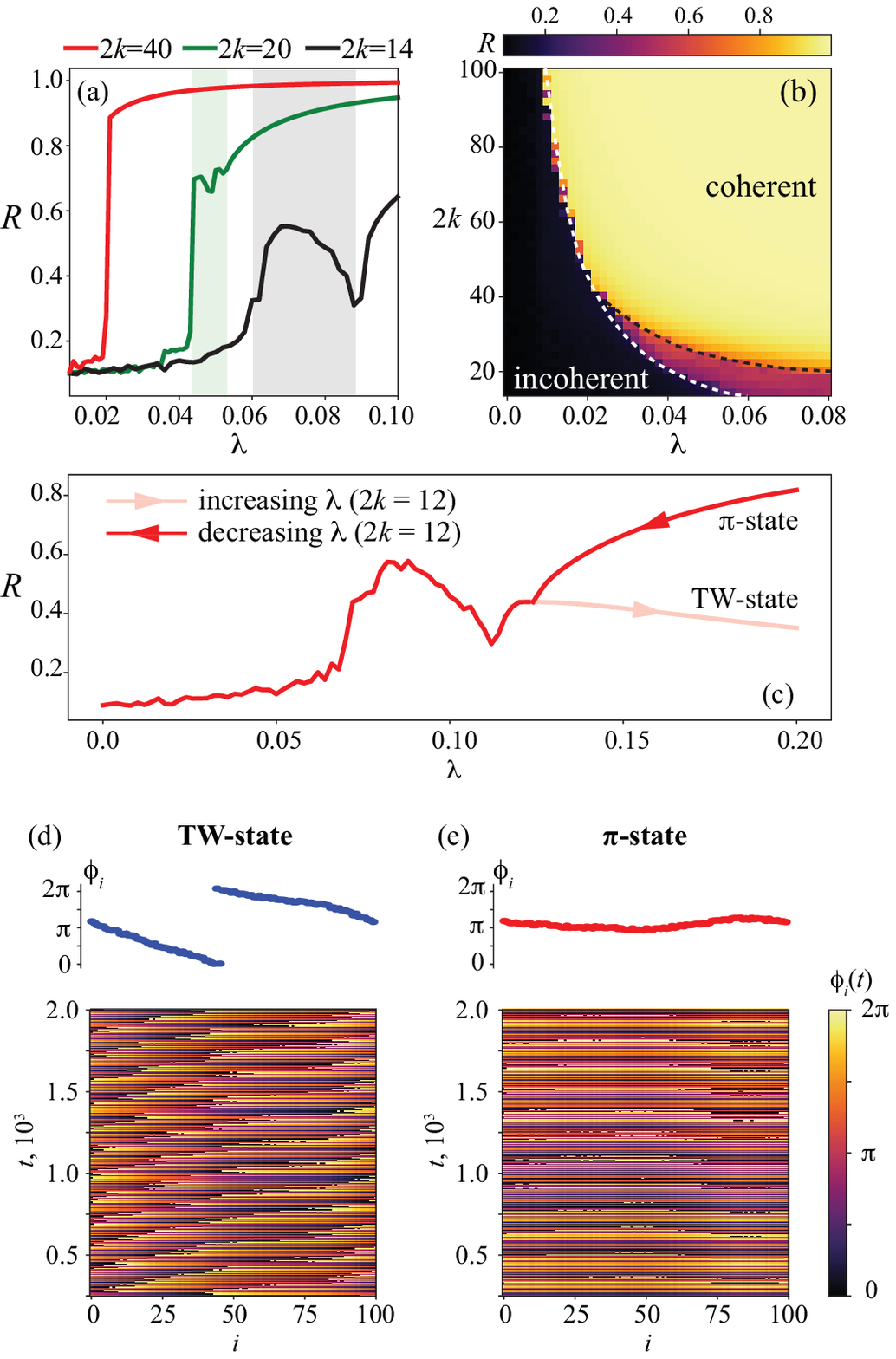}}
\caption{(a) Averaged global order parameter $R$ versus the coupling strength $\lambda$ in the non-locally coupled network of $N=100$ oscillators with $p=0.0$ and $\Delta=1.0$ for different values of nearest neighbors $2k\geq14$: $2k=14$ (black); $2k=20$ (green); $2k=40$ (red). Shadings highlight the respective areas of partially coherent chimera-like regimes. (b) Phase diagram in the $(\lambda,2k)$ parameter plane for the global order parameter $R$, color bar represents its variation. (c) $2k=12$ (exemplary illustration of the network dynamics in the case of $2k<14$). In plot (c), pink line corresponds to increasing $\lambda$ (forward transition resulting in traveling wave (TW) state and red line corresponds to decreasing $\lambda$ (backward transition resulting in $\pi$-state). Illustration of the TW (d) and frequency-locked $\pi$-state (e) for $\lambda=0.2$: instantaneous phase $\phi_i$ profiles at $t=t_{max}$ (top) and their space-time plots (bottom).}\label{fig:4}
\end{figure}

Fig.~\ref{fig:3}a shows that in the case of weak coupling strength $\lambda=0.02$, mean coupling term $\langle c_i \rangle$ remains approximately at the zero-level supported by the low values of local coherence $R_i$. After the critical transition at $\lambda=0.045$ (cf. Fig.~\ref{fig:3}b), the above described compensatory mechanism is explosively induced -- elements with $\omega_i<\omega_0$ become attractively coupled ($c_i>0$) and those with $\omega_i>\omega_0$ become repulsively coupled ($c_i<0$). Due to the uniformity of the initial natural frequency distribution, the ensemble is divided into the groups of attractive and repulsive coupling in equal proportions. Obviously, the network elements forming the attractively coupled group converge rapidly to the frequency-locked (coherent) state, i.e., $\langle f_i \rangle \rightarrow \Omega$ for all $i$ such that $c_i>0$ (cf. Fig.~\ref{fig:3}(c,d). At the same time, as also seen in Fig.~\ref{fig:3}c and Fig.~\ref{fig:3}d, repulsively coupled oscillators resist global frequency-locking at the common frequency $\Omega$. These repulsively coupled oscillators having $\langle f_i \rangle \approx \omega_0 + \Delta/2$ form a core of the separate (incoherent) cluster. Thus, the coexistence of two populations with different types of coupling determines the emergence of the chimera-like behavior in a heterogeneous Kuramoto model. Such non-homogeneity of coupling is an inevitable consequence of the frequency heterogeneity in the considered network. Finally, all elements are frequency-locked at $\lambda>0.054$ demonstrating the expected linear relation between the natural frequency $\omega_i$ and the mean coupling term $\langle c_i \rangle$ (cf. Fig.~\ref{fig:3}e).

\subsection{Influence of the network topology}

Above we have considered the formation of the chimera-like state in a heterogeneous non-locally coupled network with fixed topological properties ($p=0.0$ and $k=10$). Now, let us analyze the influence of the network topology on the transitions in the considered network model.

First, we explore how the number of the nearest neighbors $k$ affects the route to coherence in the regular non-locally coupled Kuramoto network (cf. Fig.~\ref{fig:4}a,b). Here, the previously considered network topology corresponds to a green curve. The increase of the nearest neighbors $k$ ($2k\geq40$, red curve) suppresses the emergence of a partially coherent state. As the coupling term $c_i$ summarizes the influence from all elements coupled to the $i^{th}$ one, an increase of $k$ gains the coupling term $c_i$. Besides, each element interacts with a larger group of neighboring oscillators, that counteracts the network's heterogeneity and contributes to the emergence of the first-order transition. Thus, strong interaction within the large group of elements leads to the explosive transition directly from the incoherent to a globally frequency-locked state in the absence of the intermediate partially-coherent state. On the contrary, the decrease of $k$ (black curve) promotes weaker interaction between network elements and makes it of a more local kind. These factors strengthen the influence of the network's heterogeneity, slow down the transition to coherence and support the partially coherent state in a wider range of $\lambda$.

For the values of $k$ presented in Fig.~\ref{fig:4}a, the observed transitions are reversible, i.e., the system undergoes the same transitions in both forward (increasing $\lambda$) and backward (decreasing $\lambda$) directions. Interestingly, the transition becomes irreversible with a further decrease of $k$, specifically for $2k<14$ (cf. Fig.~\ref{fig:4}c). The forward transition results in a traveling-wave (TW) solution, whose phase profile and space-time plot are presented in Fig.~\ref{fig:4}d. In turn, during the backward transition, the network converges to a more stable frequency-locked ($\pi$-state) at the high values of coupling strength (cf. Fig.~\ref{fig:4}e). We suppose, that for $2k<14$, the network topology exhibits pronounced local coupling properties, so the collective dynamics represent the interaction of locally coupled populations. Such non-homogeneity of interaction in combination with the initial heterogeneity of the network elements promote the phase lags between local interacting groups. The latter provides the convergence to a TW-solution during the forward transition under the slowly increasing coupling strength $\lambda$. During the backward transition, high coupling strength $\lambda$ forces the network to switch to a globally frequency-locked $\pi$-state. The decrease of $\lambda$ causes a smooth desynchronization of the ensemble and two solutions -- TW and $\pi$-state -- meet at the bifurcation point at $\lambda=0.124$.

\begin{figure}[!t]
\centerline{\includegraphics[scale=0.750]{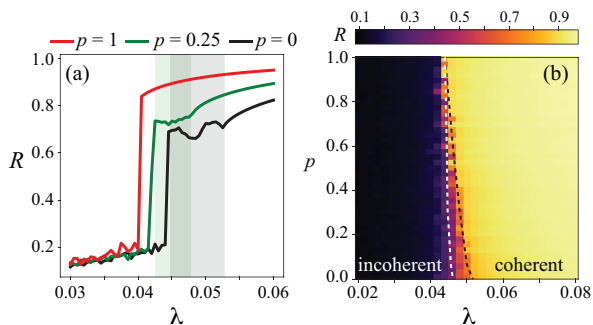}}
\caption{(a) Averaged global order parameter $R$ versus the coupling strength $\lambda$ in the non-locally coupled network of $N=100$ oscillators with $2k=20$ and $\Delta=1.0$ for different values of rewiring probability: $p = 0.0$ (black); $p = 0.25$ (green); $p=1.0$ (red). Shadings highlight the respective areas of partially coherent chimera-like regimes. (b) Phase diagram in the $(\lambda,p)$ parameter plane for the global order parameter $R$, color bar represents its variation.}\label{fig:5}
\end{figure}

\begin{figure}[!t]
	\centerline{\includegraphics[scale=0.750]{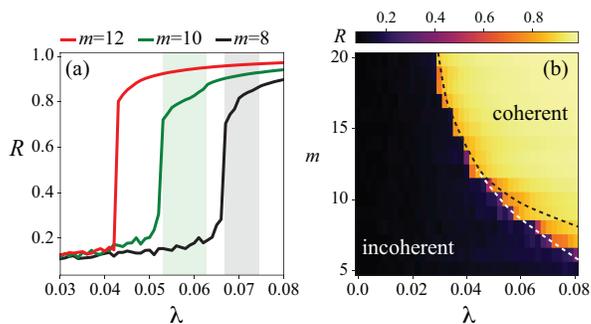}}
	\caption{(a) Averaged global order parameter $R$ versus the coupling strength $\lambda$ in the non-locally coupled network of $N=100$ oscillators with $\Delta=1.0$ for different values of the BA graph parameter $m$: $m = 8$ (black); $m=10$ (green); $m=12$ (red). Shadings highlight the respective areas of partially coherent chimera-like regimes. (b) Phase diagram in the $(\lambda,m)$ parameter plane for the global order parameter $R$, color bar represents its variation.}\label{fig:6}
\end{figure}

Finally, we consider how the structural properties of the SW and SF graphs affects the transitions of the collective behaviors. It is seen in Fig.~\ref{fig:5}, that in the case of SW topology, the increase of rewiring probability $p$ lowers the critical value of the coupling strength $\lambda_{cr}$ providing the explosive transition and smooths the area of the partially coherent state (black and green curves for $p=0.0$ and $p=0.25$, respectively). In the limit case of $p=1.0$ (completely random rewiring, red curve), the intermediate partially coherent state is suppressed by the increased network randomness resulting in the direct explosive transition from the incoherent dynamics to a frequency-locked ($\pi$-state) at $\lambda=0.0405$. Accordingly, in the SF network, the chimera-like behavior is only possible in sparsely connected graphs ($m<12$) (Fig.~\ref{fig:6}). For the dense coupling $m\geq 12$, only an explosive transition is observed.

Taken together, these results demonstrate that the detected chimera-like behavior in a heterogeneous Kuramoto model could be suppressed (i) by the increase of the neighborhood in the case of non-local coupling, (ii) by a strong rewiring in the SW network and (iii) by growing a densely coupled SF graph. We argue that these ways share a similar mechanism based on the establishment of the long-scale coupling between the network elements. Thus, the effect of initial heterogeneity of network oscillators could be annihilated by expanding the coupling area for each element, that provides the dominance of the attractive mechanisms.

\section{Conclusion}

To summarize, we have considered the transitions in a heterogeneous Kuramoto model, where the natural frequencies of its elements are chosen from a uniform distribution. Consistent with the earlier studies~\cite{laing2009chimera,laing2009dynamics}, we have demonstrated that non-homogeneity of the interacting oscillators does not ruin the emergent chimera state. Moreover, it contributes to a specific type of frequency-modulated chimera-like behavior in which frequency-locked population coexists with a non-frequency-locked one. Interestingly, the observed chimera-like pattern is not stationary -- depending on coupling strength a non-frequency-locked population appears and collapses in time either regularly or not. This is due to the origin of the chimera-like behavior. Specifically, we have shown that the interaction within the initially heterogeneous ensemble of phase oscillators leads to the splitting into the attractively and repulsively coupled groups. While the attractively coupled elements rapidly converge to a frequency-locked state, the repulsively coupled population tends to counteract the global frequency-locking, thus forming an unstable incoherent cluster.

Importantly, the uncovered chimera-like state has been observed in non-locally coupled, small-world and sparsely connected scale-free networks. On the contrary, in globally coupled networks, networks with completely random rewiring and densely connected scale-free networks, the ensemble undergoes the direct transition from the incoherent state to a global frequency-locking. We conclude that in the latter networks, the emergence of large-scale connections contributes to the dominance of the attractive coupling by influencing excitatory on a larger group of oscillators. We also hypothesize that this mechanism could be used in the real-world networks exhibiting strong rewiring of links, for example, brain neural networks, to overcome the inherent heterogeneity of its elements and suppress partially coherent states.

\begin{acknowledgments}
    	This work has been supported by the Russian Foundation for Basic Research (Grant No. 19-52-45026) and the Department of Science and Technology, Government of India (Project no. INT/RUS/RFBR/360).
\end{acknowledgments}

\section{Data Availability}

All numerical experiments with a heterogeneous Kuramoto model are described in the paper and can be reproduced without additional information. 

\section*{References}
%\bibliography{cas-refs}

\bibliographystyle{aip}

%\bibliography{cas-refs}


\end{document}